\documentclass[aps,prl,twocolumn,superscriptaddress,showpacs,preprintnumbers,amsmath,amssymb]{revtex4-1}

\usepackage{graphicx}
\usepackage{dcolumn}
\usepackage{lineno}
\usepackage{epstopdf}
\usepackage{amsmath}
\usepackage{xcolor}
\RequirePackage{xspace}
\usepackage{relsize}
\def\babar{\mbox{\slshape B\kern-0.1em{\smaller A}\kern-0.1em B\kern-0.1em{\smaller A\kern-0.2em R}}}
\def\Bbar  {\ensuremath{\overline B}\xspace}
\def\ccbar  {\ensuremath{c\overline c}\xspace}
\def\qqbar  {\ensuremath{q\overline q}\xspace}
\def\KS    {\ensuremath{K^0_{\scriptscriptstyle S}}\xspace} 
\def\KL    {\ensuremath{K^0_{\scriptscriptstyle L}}\xspace} 

\def\Dstarp{\ensuremath{D^{*+}}\xspace}
\mathchardef\Upsilon="7107
\def\Y#1S{\ensuremath{\Upsilon{(#1S)}}\xspace}
\def\cm    {\ensuremath{{\rm \,cm}}\xspace}
\def\invfb {\ensuremath{\mbox{\,fb}^{-1}}\xspace}
\def\order {{\ensuremath{\cal O}}\xspace}

\def\Neps  {\ensuremath{N/\varepsilon}\xspace}
\def\etal  {{\it et~al.}}
\newcommand{\gev}{\ensuremath{\rm{\,Ge\kern -0.1em V}}\xspace}
\newcommand{\mev}{\ensuremath{\rm{\,Me\kern -0.1em V}}\xspace}
\newcommand{\gevc}{\ensuremath{{\rm{\,Ge\kern -0.1em V\!/}c}}\xspace}
\newcommand{\mevc}{\ensuremath{{\rm{\,Me\kern -0.1em V\!/}c}}\xspace}
\newcommand{\gevcc}{\ensuremath{{\rm{\,Ge\kern -0.1em V\!/}c^2}}\xspace}
\newcommand{\mevcc}{\ensuremath{{\rm{\,Me\kern -0.1em V\!/}c^2}}\xspace}

\graphicspath{{ps}}

\renewcommand{\arraystretch}{1.1}

\begin{document}

\preprint{\vbox{ \hbox{ }
                        \hbox{Belle Preprint {\it 2015-18}}
                        \hbox{KEK Preprint {\it 2015-51}}
}}

\title{\quad\\[0.5cm]
Search for the rare decay {\boldmath $D^0\to\gamma\gamma$} at Belle}

\noaffiliation
\affiliation{Aligarh Muslim University, Aligarh 202002}
\affiliation{University of the Basque Country UPV/EHU, 48080 Bilbao}
\affiliation{University of Bonn, 53115 Bonn}
\affiliation{Budker Institute of Nuclear Physics SB RAS, Novosibirsk 630090}
\affiliation{Faculty of Mathematics and Physics, Charles University, 121 16 Prague}
\affiliation{Chonnam National University, Kwangju 660-701}
\affiliation{University of Cincinnati, Cincinnati, Ohio 45221}
\affiliation{Deutsches Elektronen--Synchrotron, 22607 Hamburg}
\affiliation{University of Florida, Gainesville, Florida 32611}
\affiliation{Justus-Liebig-Universit\"at Gie\ss{}en, 35392 Gie\ss{}en}
\affiliation{Gifu University, Gifu 501-1193}
\affiliation{SOKENDAI (The Graduate University for Advanced Studies), Hayama 240-0193}
\affiliation{Hanyang University, Seoul 133-791}
\affiliation{University of Hawaii, Honolulu, Hawaii 96822}
\affiliation{High Energy Accelerator Research Organization (KEK), Tsukuba 305-0801}
\affiliation{IKERBASQUE, Basque Foundation for Science, 48013 Bilbao}
\affiliation{Indian Institute of Technology Bhubaneswar, Satya Nagar 751007}
\affiliation{Indian Institute of Technology Madras, Chennai 600036}
\affiliation{Indiana University, Bloomington, Indiana 47408}
\affiliation{Institute of High Energy Physics, Chinese Academy of Sciences, Beijing 100049}
\affiliation{Institute of High Energy Physics, Vienna 1050}
\affiliation{INFN - Sezione di Torino, 10125 Torino}
\affiliation{Institute for Theoretical and Experimental Physics, Moscow 117218}
\affiliation{J. Stefan Institute, 1000 Ljubljana}
\affiliation{Kanagawa University, Yokohama 221-8686}
\affiliation{Institut f\"ur Experimentelle Kernphysik, Karlsruher Institut f\"ur Technologie, 76131 Karlsruhe}
\affiliation{Kennesaw State University, Kennesaw GA 30144}
\affiliation{King Abdulaziz City for Science and Technology, Riyadh 11442}
\affiliation{Korea Institute of Science and Technology Information, Daejeon 305-806}
\affiliation{Korea University, Seoul 136-713}
\affiliation{Kyoto University, Kyoto 606-8502}
\affiliation{Kyungpook National University, Daegu 702-701}
\affiliation{\'Ecole Polytechnique F\'ed\'erale de Lausanne (EPFL), Lausanne 1015}
\affiliation{Faculty of Mathematics and Physics, University of Ljubljana, 1000 Ljubljana}
\affiliation{Ludwig Maximilians University, 80539 Munich}
\affiliation{Luther College, Decorah, Iowa 52101}
\affiliation{University of Maribor, 2000 Maribor}
\affiliation{Max-Planck-Institut f\"ur Physik, 80805 M\"unchen}
\affiliation{School of Physics, University of Melbourne, Victoria 3010}
\affiliation{Moscow Physical Engineering Institute, Moscow 115409}
\affiliation{Moscow Institute of Physics and Technology, Moscow Region 141700}
\affiliation{Graduate School of Science, Nagoya University, Nagoya 464-8602}
\affiliation{Kobayashi-Maskawa Institute, Nagoya University, Nagoya 464-8602}
\affiliation{Nara Women's University, Nara 630-8506}
\affiliation{National Central University, Chung-li 32054}
\affiliation{National United University, Miao Li 36003}
\affiliation{Department of Physics, National Taiwan University, Taipei 10617}
\affiliation{H. Niewodniczanski Institute of Nuclear Physics, Krakow 31-342}
\affiliation{Niigata University, Niigata 950-2181}
\affiliation{University of Nova Gorica, 5000 Nova Gorica}
\affiliation{Novosibirsk State University, Novosibirsk 630090}
\affiliation{Osaka City University, Osaka 558-8585}
\affiliation{Pacific Northwest National Laboratory, Richland, Washington 99352}
\affiliation{Panjab University, Chandigarh 160014}
\affiliation{Peking University, Beijing 100871}
\affiliation{University of Pittsburgh, Pittsburgh, Pennsylvania 15260}
\affiliation{Punjab Agricultural University, Ludhiana 141004}
\affiliation{University of Science and Technology of China, Hefei 230026}
\affiliation{Seoul National University, Seoul 151-742}
\affiliation{Soongsil University, Seoul 156-743}
\affiliation{University of South Carolina, Columbia, South Carolina 29208}
\affiliation{Sungkyunkwan University, Suwon 440-746}
\affiliation{School of Physics, University of Sydney, NSW 2006}
\affiliation{Department of Physics, Faculty of Science, University of Tabuk, Tabuk 71451}
\affiliation{Tata Institute of Fundamental Research, Mumbai 400005}
\affiliation{Excellence Cluster Universe, Technische Universit\"at M\"unchen, 85748 Garching}
\affiliation{Department of Physics, Technische Universit\"at M\"unchen, 85748 Garching}
\affiliation{Toho University, Funabashi 274-8510}
\affiliation{Department of Physics, Tohoku University, Sendai 980-8578}
\affiliation{Earthquake Research Institute, University of Tokyo, Tokyo 113-0032}
\affiliation{Department of Physics, University of Tokyo, Tokyo 113-0033}
\affiliation{Tokyo Institute of Technology, Tokyo 152-8550}
\affiliation{Tokyo Metropolitan University, Tokyo 192-0397}
\affiliation{University of Torino, 10124 Torino}
\affiliation{Utkal University, Bhubaneswar 751004}
\affiliation{CNP, Virginia Polytechnic Institute and State University, Blacksburg, Virginia 24061}
\affiliation{Wayne State University, Detroit, Michigan 48202}
\affiliation{Yamagata University, Yamagata 990-8560}
\affiliation{Yonsei University, Seoul 120-749}
  \author{N.~K.~Nisar}\affiliation{Tata Institute of Fundamental Research, Mumbai 400005}\affiliation{Aligarh Muslim University, Aligarh 202002} 
  \author{G.~B.~Mohanty}\affiliation{Tata Institute of Fundamental Research, Mumbai 400005} 
  \author{K.~Trabelsi}\affiliation{High Energy Accelerator Research Organization (KEK), Tsukuba 305-0801}\affiliation{SOKENDAI (The Graduate University for Advanced Studies), Hayama 240-0193} 
  \author{T.~Aziz}\affiliation{Tata Institute of Fundamental Research, Mumbai 400005} 
  \author{A.~Abdesselam}\affiliation{Department of Physics, Faculty of Science, University of Tabuk, Tabuk 71451} 
  \author{I.~Adachi}\affiliation{High Energy Accelerator Research Organization (KEK), Tsukuba 305-0801}\affiliation{SOKENDAI (The Graduate University for Advanced Studies), Hayama 240-0193} 
  \author{H.~Aihara}\affiliation{Department of Physics, University of Tokyo, Tokyo 113-0033} 
  \author{D.~M.~Asner}\affiliation{Pacific Northwest National Laboratory, Richland, Washington 99352} 
  \author{V.~Aulchenko}\affiliation{Budker Institute of Nuclear Physics SB RAS, Novosibirsk 630090}\affiliation{Novosibirsk State University, Novosibirsk 630090} 
  \author{T.~Aushev}\affiliation{Moscow Institute of Physics and Technology, Moscow Region 141700}\affiliation{Institute for Theoretical and Experimental Physics, Moscow 117218} 
% ...
\collaboration{The Belle Collaboration}

\begin{abstract}

We search for the rare radiative decay $D^0\to\gamma\gamma$ using a
data sample with an integrated luminosity of $832\invfb$ recorded by
the Belle detector at the KEKB $e^+e^-$ asymmetric-energy collider.
We find no statistically significant signal and set an upper limit on
the branching fraction of ${\cal B}(D^0\to\gamma\gamma)<8.5\times10^{-7}$
at $90\%$ confidence level. This is the most restrictive limit on the
decay channel to date.

\end{abstract}

\maketitle
{\renewcommand{\thefootnote}{\fnsymbol{footnote}}}
\setcounter{footnote}{0}


Flavor-changing neutral current (FCNC) processes are forbidden
at tree level in the standard model (SM), although they can occur
at higher orders. In contrast, there are several new physics (NP)
models that allow FCNC even at tree level and can substantially
enhance the branching fractions of the related decay processes.
The decay of the neutral charm meson to two photons, $D^0\to
\gamma\gamma$, is one such example that constitutes a sensitive
NP probe. Mediated by a $c\to u\gamma\gamma$ transition, the
amplitude for this rare decay has very small [$\order(10^{-11})$]
short-distance contributions~\cite{theo1,theo2,theo3} due to the
small size of the bottom-quark mass relative to the weak scale
and a low value of the quark-mixing matrix~\cite{ckm} element $V_{ub}$.
However, it is expected to have large long-distance contributions
from intermediate vector mesons. Theory calculations based on vector
meson dominance yield a decay branching fraction in the range
$(1$--$3)\times10^{-8}$~\cite{theo2,theo3}. Using the framework of
the minimal supersymmetric standard model (MSSM), the authors of
Ref.~\cite{theo4} have predicted that the exchange of gluinos ---
the supersymmetric partners of gluons --- can enhance the branching
fraction up to $6\times10^{-6}$. Furthermore, it has been suggested
that, by measuring $D^0\to\gamma\gamma$, one would be able to better
identify potential NP contributions to $D^0\to\mu^+\mu^-$~\cite{theo5}.

Searches for the $D^0\to\gamma\gamma$ decay have been previously conducted
by the CLEO~\cite{cleo_res} and $\babar$~\cite{babar_res} experiments using
$e^+e^-$ collision data recorded at the $\Y4S$ resonance and recently by
BESIII~\cite{bes3_res} based on data collected near the open-charm threshold. 
The most stringent upper limit on the branching fraction is that set by
$\babar$: $2.2\times10^{-6}$ at $90\%$ confidence level (CL).

We report herein a search for $D^0\to\gamma\gamma$~\cite{conjugate}
using a data sample of $832\invfb$ collected near the $\Y4S$
and $\Y5S$ resonances with the Belle detector~\cite{belle} at
the KEKB asymmetric-energy $e^+e^-$ collider~\cite{kekb}. The
detector elements most relevant for the study are a tracking
device comprising a silicon vertex detector and a central
drift chamber (CDC); a particle identification system that
consists of a barrel-like arrangement of time-of-flight counters
(TOF) and an array of aerogel threshold Cherenkov counters (ACC);
and a CsI(Tl) crystal-based electromagnetic calorimeter (ECL).
All these components are located inside a superconducting solenoid
coil that provides a $1.5$\,T magnetic field. In addition to data,
we use Monte Carlo (MC) simulated events to devise selection
criteria and study possible backgrounds. The relative size
of the $\Y4S$ and $\Y5S$ MC samples is determined according
to the integrated luminosity of the corresponding data.

To reduce large `combinatorial' backgrounds arising from random
photon combinations, we require that the $D^0$ be produced in the
decay $\Dstarp\to D^0\pi^+$. The $\Dstarp$ mesons mostly originate
from the $e^+e^-\to\ccbar$ process via hadronization, where the
inclusive yield has a large uncertainty of $12.5\%$~\cite{PDG}. To
avoid this uncertainty, we measure the $D^0\to\gamma\gamma$ branching
fraction with respect to the well-measured mode $D^0\to\KS\pi^0$
using the following relation:
\begin{equation}
{\cal B}(D^0\to\gamma\gamma)=\frac{(\Neps)_{D^0\to\gamma\gamma}}{(\Neps)_{D^0\to\KS\pi^0}}
\times{\cal B}(D^0\to\KS\pi^0).
\label{eq_bf}
\end{equation}
Here, $N$ and $\varepsilon$ are the signal yields and detection
efficiencies, respectively, of the reconstructed channels and
${\cal B}(D^0\to\KS\pi^0)$ is the world-average branching fraction
for $D^0\to\KS\pi^0$~\cite{PDG}. Also, systematic uncertainties
common to both the signal and normalization channels cancel in
this measurement.

We identify photon candidates as localized energy deposits
(``clusters") in the ECL without any matched charged track
in the CDC and having an energy greater than $200\mev$.
Track candidates are required to have an impact parameter
with respect to the interaction point (IP) of less than $1\cm$
in the transverse plane and less than $3\cm$ along the $+z$
axis (opposite the $e^+$ beam). Charged pions are
distinguished from kaons~\cite{pid} using specific ionization
in the CDC, time-of-flight information from the TOF, and the
number of photoelectrons from the ACC. The pion identification
efficiency is above $95\%$ while the probability to misidentify
a kaon as a pion is below $5\%$. Candidate $\KS$ mesons
are reconstructed from pairs of oppositely charged tracks (pion
mass assumed) having a reconstructed mass within $9\mevcc$
of the nominal $\KS$ mass~\cite{PDG}. We reconstruct $\pi^0$
candidates from diphoton pairs with an invariant mass in the
range $(110$--$160)\mevcc$. More details on the  $\KS$ and
$\pi^0$ reconstruction can be found in Ref.~\cite{Acp-pi0pi0}.

We reconstruct a $D^0$ candidate from two energetic photons. The
$D^0$ candidate is then combined with a low-momentum (``slow'')
pion, $\pi^+_{s}$, to form a $\Dstarp$. We use two kinematic 
variables to identify signal: the reconstructed invariant mass of
the $D^0$ candidate, $M(\gamma\gamma)$, and the difference
between the reconstructed masses of the $\Dstarp$ and $D^0$
candidates, $\Delta M$. To improve the $\Delta M$ resolution,
the $\pi^+_{s}$ track is constrained to originate from the
IP. The $D^0$ candidate in the normalization channel is
formed by combining a $\KS$ with a $\pi^0$ candidate. The
signal peaks near the nominal $D^0$ mass~\cite{PDG} and
at $145\mevcc$ in the $M(\gamma\gamma)$ and $\Delta M$
distributions, respectively. We select candidate events that
satisfy the criteria $1.7\gevcc<M(\gamma\gamma)<2.0\gevcc$ and
$140\mevcc<\Delta M<160\mevcc$. Furthermore, we define a smaller
signal region as $1.711\gevcc<M(\gamma\gamma)<1.931\gevcc$ and
$143.4\mevcc<\Delta M<147.7\mevcc$ [$\pm3\sigma$ windows around
the means of the $M(\gamma\gamma)$ and $\Delta M$ distributions],
and a sideband as $1.95\gevcc<M(\gamma\gamma)<2.00\gevcc$ and
$150\mevcc<\Delta M<160\mevcc$.

We study various backgrounds to the $D^0\to\gamma\gamma$ signal.
These can be broadly classified into three categories: peaking,
QED, and combinatorial. The first is from specific physics
processes such as $D^0\to\pi^0\pi^0$, $D^0\to\eta\pi^0$,
$D^0\to\eta\eta$, $D^0\to\KS(\pi^0\pi^0)\pi^0$, and $D^0\to\KL\pi^0$.
These processes can be misidentified as signal in two different ways.
One possibility is from a pair of high-energy photons, either one or both
coming from a $\pi^0$ or $\eta$ decay. This is suppressed by pairing
each photon candidate of $D^0\to\gamma\gamma$ with all other photons
in the event and applying criteria on the resulting probabilities,
${\cal P}(\pi^0)$ and ${\cal P}(\eta)$~\cite{pi0veto}. The second
possibility is due to merged clusters in the ECL owing to a small
opening angle between two photons from a high-momentum $\pi^0$ or
$\eta$ decay. Such clusters are wider in the lateral dimension and
are rejected by requiring that the energy deposited in the $3\times3$
array of crystals centered on the crystal with the highest energy
exceeds $85\%$ of the energy deposited in the corresponding $5\times
5$ array of crystals. In the case of $D^0\to\KL\pi^0$,
misidentification of the $\KL$ as a $\gamma$ candidate contributes
to the background. The peaking background exhibits a signal-like peak
in the $\Delta M$ distribution but peaks at a lower $M(\gamma\gamma)$
value due to particles missing from the reconstruction.

The QED background arises from out-of-time $e^+e^-\to\gamma\gamma$
and $e^+e^-\to e^+e^-(\gamma)$ events. To suppress its contribution, we
retain only those events in which the number of charged tracks and photon
candidates each exceeds four. We also require that the timing of ECL
clusters for photon candidates be consistent with the beam collision
time identified at the trigger level. With these criteria, we find a
negligible contribution from the QED background.

To suppress remaining backgrounds, we apply selection requirements on
the following variables: the momentum of the $D^*$ candidate in the $e^+e^-$
center-of-mass frame, $p^*(D^*)$;  the energy asymmetry between the two
photons, $A_E=(E_{\gamma 1}-E_{\gamma 2})/(E_{\gamma 1}+E_{\gamma 2})$,
where $E_{\gamma 1\,(2)}$ is the energy  of the higher\,(lower) energy
photon; $E_{\gamma2}$; ${\cal P}(\pi^0)$; and ${\cal P}(\eta)$. The
requirements are determined using an optimization procedure~\cite{g_punzi}
with a figure-of-merit given by
\begin{equation}
{\rm FOM}=\frac{\varepsilon(t)}{\sqrt{N_B(t)}},
\label{eq_fom}
\end{equation}
where $t$ is the value of the selection criterion and $\varepsilon$ and $N_B$
are, respectively, the detection efficiency and the number of background
events expected in the signal region. As the contribution of $\sqrt{N_B(t)}$
is overwhelming, we neglect the $a/2$ term in the denominator of the original
FOM expression in Ref.~\cite{g_punzi}, where $a$ is the desired CL in terms
of standard deviations. We use signal MC events to estimate $\varepsilon$
and a blended MC sample of $e^+e^-\to\qqbar\,(q=u,d,s,c)$ and $B\Bbar$ events
to calculate $N_B$. Some of the variables, notably $E_{\gamma 2}$ and $A_E$,
can have significant correlations among themselves. We take this effect into
account via simultaneous optimizations of the criteria in these variables.

To incorporate possible differences between data and simulations in
the optimization procedure, we multiply $N_B$ in Eq.~(\ref{eq_fom})
by a correction factor estimated by comparing data and MC events
in the sideband. The factor could be as large as $1.6$, depending
on the variable. The selection criteria obtained are $p^*(D^*)>
2.9\gevc$, $E_{\gamma 2}>900\mev$, $A_E<0.5$, ${\cal P}(\pi^0)
<0.15$, and no requirement on ${\cal P}(\eta)$. We have verified
that the same set of criteria can be applied to the combined sample
of $\Y4S$ and $\Y5S$ data with the level of backgrounds being
proportional to the respective luminosity.

The efficiency for signal events to pass the above selection criteria is
$(7.34\pm0.05)\%$. About $3\%$ of the events selected in the signal MC
sample have multiple $\Dstarp$ candidates. In these cases, we retain the
one with the smallest value of the $\pi^+_s$ impact parameter with respect
to the IP in the transverse plane. If there is a unique $\pi^+_s$ but multiple
$D^0$ candidates, the first one is arbitarily chosen. From simulations, we
find that this criterion identifies the correct $\Dstarp$ decay in $70\%$
of the cases.

Event candidates for the normalization channel $D^0\to\KS\pi^0$ are
selected with criteria similar to those of the signal for $\pi^{+}_{s}$,
$p^*(D^*)$, $\Delta M$, and the invariant mass of the $D^0$ candidate.
For the criteria that differ from those of the signal selection, we rely
on Ref.~\cite{Acp-pi0pi0}. With these requirements, the detection efficiency
is found to be $(7.18\pm0.05)\%$.

To extract the signal yield, we perform an unbinned extended maximum
likelihood fit to the two-dimensional (2D) distributions of $M(\gamma
\gamma)$ and $\Delta M$. For the signal and combinatorial-background
component, the correlation between the two observables is negligible.
Thus we define a combined probability density function (PDF) for each
component, indexed by $j$, as
\begin{equation}
{\cal P}_j\equiv{\cal P}_j[M(\gamma\gamma)]\,{\cal P}_j[\Delta M].
\label{eq_pdf}
\end{equation}
The signal shape in both $M(\gamma\gamma)$ and $\Delta M$ is described
by a sum of a Gaussian and an asymmetric Gaussian function of common
mean. The combinatorial background shape in $M(\gamma\gamma)$
is modeled by a third-order polynomial and in $\Delta M$ by a threshold
function $(x-m_{\pi^+})^{\alpha}\exp[-\beta(x-m_{\pi^+})]$, where $\alpha$
and $\beta$ are two shape parameters and $m_{\pi^+}$ is the nominal
charged pion mass~\cite{PDG}. For the peaking background, there
is a significant correlation between $\Delta M$ and $M(\gamma\gamma)$,
which we account for via a joint PDF ${\cal P}[\Delta M|M(\gamma\gamma)]$. 
This background component is described by a single Gaussian function in
$M(\gamma\gamma)$ and, just as for the signal, the sum of a Gaussian and
an asymmetric Gaussian function in $\Delta M$. To include the correlation,
we parametrize the width of the Gaussian part of $\Delta M$ as $\sigma=
\sigma_0(1+k[M(\gamma\gamma)-m_{D^{0}}]^2)$, where $\sigma_0$ is the
uncorrelated value, $k$ is the correlation coefficient, and $m_{D^{0}}$ is
the world-average $D^0$ mass~\cite{PDG}. The two widths of the asymmetric
Gaussian component are scaled from $\sigma$. All combinatorial PDF
parameters are determined from the data fit, while those for signal and
peaking background are fixed to the corresponding MC values. 

To take possible data--MC difference into account for the signal PDFs, we use a
sample of $D^0\to\phi(K^+K^-)\gamma$, from which we extract the shift in mean
values and the ratio of widths between data and simulations as calibration factors.
The control mode suffers from significant contamination from $D^0\to\phi(K^+K^-)
\pi^0$. To better discriminate signal from this background, we include the helicity
angle in the fit, which is defined as the angle between the $K^+$ momentum
and the negative of the $D^0$ momentum in the $\phi$ rest frame. As we wish
to apply correction factors obtained from $D^0\to\phi\gamma$, which contains
one photon, to the signal channel with two photons in the final state, we shift
the MC $M(\gamma\gamma)$ mean value by twice its correction and multiply the
width by the square of the corresponding correction factor. On the other hand,
the $\Delta M$ resolution is dominated by the momentum measurement of
$\pi^+_s$, for which there is no difference between the signal and control
channel. Therefore, the $\Delta M$ corrections are applied without any change.

To calibrate the peaking background shape in $M(\gamma\gamma)$, we
compare data and MC distributions in a sample of $D^0\to\pi^0\pi^0$ that
is partially reconstructed using the higher-energy photons from each $\pi^0$
decay. The $\Delta M$ correction factors are obtained using a sample of
candidates in data and MC events for the forbidden decay $D^0\to\KS\gamma$,
where the selected candidates are mostly due to partially reconstructed $D^0
\to\KS\pi^0$ decays.

We apply the fit to simulated MC samples and obtain yields for the
three event categories that are consistent with their input values.
Furthermore, we check the stability and error coverage of the fit
by applying it to an ensemble of pseudo-experiments where events are
drawn from the PDF shapes for all three event categories as described
above. The exercise is repeated for various possible signal yields
ranging from $0$ to $100$. We find a negligible bias on the fitted signal
yield and the latter consistent with the input value within uncertainties.

\begin{figure}[htbp]
\begin{center}
 \begin{tabular}{cc}
   \includegraphics[width=0.49\textwidth]{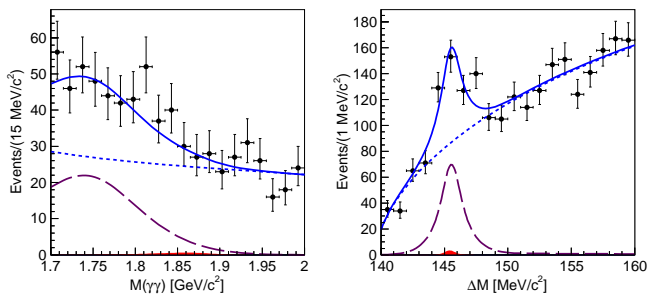}
\end{tabular}
\caption{Projections of candidate events onto the $M(\gamma\gamma)$ (left)
and $\Delta M$ (right) distributions, applying a signal-region criterion
on the other variable. Points with error bars are the data, blue solid curves
are the results of the fit, blue dotted curves represent the combinatorial
background, magenta dashed curves are the peaking background, and red
filled histograms show the signal component.}
\label{fit_res}
\end{center}
\end{figure}

Applying the 2D fit described above to the $3148$ candidate events,
we find $4\pm15$ signal, $210\pm32$ peaking background and $2934\pm59$
combinatorial background events. Figure~\ref{fit_res} shows the
results of the fit. In the absence of a statistically significant signal,
we derive an upper limit at $90\%$ CL on the signal yield
($N^{90\%}_{\rm UL}$) following a frequentist method~\cite{ul_method}
using an ensemble of pseudo-experiments. For a given signal yield,
we generate $5000$ sets of signal and background events according
to their PDFs, and perform the fit. The CL is obtained by calculating
the fraction of samples that gives a fit yield larger than that observed
in data ($4$ events). The systematic uncertainty (described below) is
accounted for in the limit calculation by smearing the fit yield. We
obtain $N^{90\%}_{\rm UL}$ to be $25$ events.

As this is a relative measurement, most of the systematic uncertainties
common between the signal and normalization channels cancel. However,
some residual systematics remain. We estimate their contributions by
varying the selection criteria that do not necessarily factor out.
These include $E_{\gamma 2}$, $A_E$, and ${\cal P}(\pi^0)$. For
$E_{\gamma 2}$ we estimate $\Neps$ with and without any requirement
on the photon energy in the $D^0\to\phi\gamma$ control sample. The
change with respect to the nominal value is taken as the corresponding
systematic error. The uncertainty due to the ${\cal P}(\pi^0)$ requirement
is calculated in the same control sample by comparing the nominal yield
with the one obtained with a substantially relaxed criterion [${\cal P}
(\pi^0)<0.7$]. We double the above systematic uncertainties, as our signal
has two photons. Since we do not have a proper control sample for $A_E$,
we fit to the data without this requirement and take the resulting change
in the upper limit as the systematic error.

Another source of systematics is due to the calibration factors
applied to MC-determined PDF shapes for the fit to data. In case of
signal, we repeat the fit by varying the PDF shapes in accordance
with the uncertainties obtained in the $D^0\to\phi\gamma$ control
channel and take the change in the signal yield as the systematic
error. To estimate the PDF shape uncertainty due to the peaking
background, similar exercises are also performed by changing the
corresponding calibration factors by $\pm1\sigma$.

Finally, there is a systematic uncertainty in the efficiencies for photon
detection, $\KS$, and $\pi^0$ reconstruction. The systematic error due to
photon detection is about $2.2\%$ for $E_{\gamma}=1\gev$~\cite{xb_ref}.
With two energetic photons in the signal final state, we assign a $4.4\%$
uncertainty. The uncertainty associated with $\KS$ reconstruction is
estimated with a sample of $\Dstarp\to D^0\pi^{+}_{s},D^0\to\KS(\pi^+\pi^-)
\pi^+\pi^-$ decays and is $0.7\%$. We obtain the systematic error due to
$\pi^0$ reconstruction ($4.0\%$) by comparing data--MC differences of the
yield ratio between $\eta\to\pi^0\pi^0\pi^0$ and $\eta\to\pi^+\pi^-\pi^0$. 
The last error is that on the branching fraction of the normalization
channel $D^0\to\KS\pi^0$~\cite{PDG}. Table~\ref{syst} summarizes all
systematic sources along with their contributions.

\begin{figure}[htbp]
\begin{center}
 \begin{tabular}{cc}
   \includegraphics[width=0.49\textwidth]{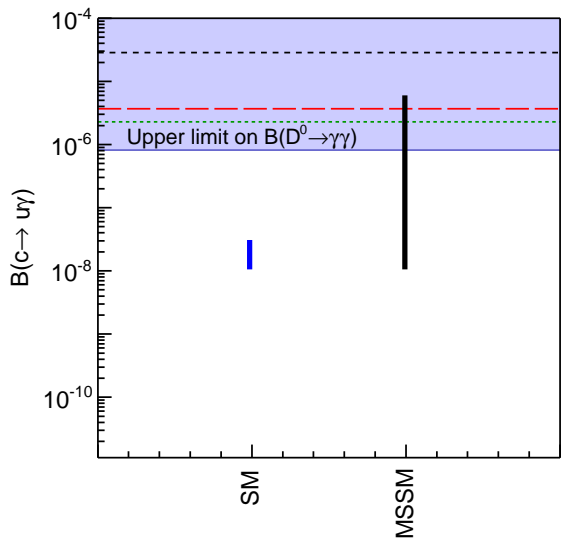}
\end{tabular}
\caption{Ranges of the $c\to u\gamma$ branching fraction predicted in
the SM and MSSM~\cite{theo4} are compared with our obtained upper
limit on ${\cal B}(D^0\to\gamma\gamma)$, shown by the purple solid
line. The limits from \babar~\cite{babar_res}, BESIII~\cite{bes3_res}, and
CLEO~\cite{cleo_res} are indicated by the green dotted, red long-dashed,
and black dashed lines, respectively.}
\label{ul_comp}
\end{center}
\end{figure}

The 2D fit is then applied to the normalization channel of $D^0\to
\KS\pi^0$, using the same signal and background models as for
$D^0\to\gamma\gamma$. All signal shape parameters are floated
during the fit. We find a signal yield of $343\,050\pm673$ events.
Using the above information in Eq.~(\ref{eq_bf}), we obtain a
$90\%$ CL upper limit on the branching fraction of ${\cal B}(D^0
\to\gamma\gamma)<8.5\times10^{-7}$. In Fig.~\ref{ul_comp}, we
compare our upper limit with those obtained by CLEO, BESIII and
\babar\ as well as with the $c\to u\gamma$ branching fractions
expected in the SM and MSSM~\cite{theo4}.

\begin{table}[tbh]
\caption{Summary of systematic uncertainties for $D^0\to\gamma\gamma$.}
\label{syst}
\renewcommand{\arraystretch}{1.0}
\begin{ruledtabular}
\begin{tabular}{l|c}
Source  &  Contribution \\
\hline
Cut variation  & $\pm6.8\%$\\
PDF shape  & $^{+4.0}_{-2.4}$ events\\
Photon detection & $\pm4.4\%$ \\
$\KS$ reconstruction & $\pm0.7\%$\\
$\pi^0$ identification & $\pm4.0\%$ \\
${\cal B}(D^0\to\KS\pi^0)$ & $\pm3.3\%$\\
\end{tabular}
\end{ruledtabular}
\end{table}

In summary, we search for the rare decay $D^0\to\gamma\gamma$
using the full data sample recorded by the Belle experiment at
or above the $\Y4S$ resonance. In the absence of a statistically
significant signal, a $90\%$ CL upper limit is set on its branching
fraction of $8.5\times10^{-7}$. Our result constitutes the most
restrictive limit on $D^0\to\gamma\gamma$ to date and can be used
to constrain NP parameter spaces. This FCNC decay will be probed
further at the next-generation Belle II experiment~\cite{belle2}.
\\

We thank the KEKB group for excellent operation of the
accelerator; the KEK cryogenics group for efficient solenoid
operations; and the KEK computer group, the NII, and 
PNNL/EMSL for valuable computing and SINET4 network support.  
We acknowledge support from MEXT, JSPS and Nagoya's TLPRC (Japan);
ARC (Australia); FWF (Austria); NSFC and CCEPP (China); 
MSMT (Czechia); CZF, DFG, and VS (Germany); DST (India); INFN (Italy); 
MOE, MSIP, NRF, BK21Plus, WCU and RSRI  (Korea);
MNiSW and NCN (Poland); MES and RFAAE (Russia); ARRS (Slovenia);
IKERBASQUE and UPV/EHU (Spain); 
SNSF (Switzerland); NSC and MOE (Taiwan); and DOE and NSF (USA).


\begin{thebibliography}{99}
\bibitem{conjugate}
 Throughout this paper, the charge-conjugate decay modes are always
 implied unless stated otherwise.
\end{thebibliography}
\end{document}